# Effect of lattice distortions on the electron and thermal transport properties of transparent oxide semiconductor $Ba_{1-x}Sr_xSnO_3$ solid solution films

Hai Jun Cho,[1, 2, a)] Koichi Sato,[2] Mian Wei,[2] Gowoon Kim,[2] and Hiromichi Ohta[1, 2, a)]

AFFILIATIONS

[1] Research Institute for Electronic Science, Hokkaido University, N20W10, Sapporo 001−0020, Japan

[2] Graduate School of Information Science and Engineering, Hokkaido University, N14W9, Kita, Sapporo 060−0814, Japan

[a)]Authors to whom correspondence should be addressed: joon@es.hokudai.ac.jp and hiromichi.ohta@es.hokudai.ac.jp

ABSTRACT

La-doped $A$SnO$_3$ ($A$ = Ba, Sr) have great potential as advanced transparent oxide semiconductors due to their large optical bandgap and relatively high electron mobility. The bandgap of Ba$_{1-x}$Sr$_x$SnO$_3$ solid solution increases from 3.2 eV (BaSnO$_3$) to 4.6 eV (SrSnO$_3$) with $x$. However, the increase in the bandgap is accompanied by reductions in the electrical conductivity. The versatility in the changes in the electrical properties are not trivial, and the property optimization has been challenging. Here we propose a simple metric for quantifying the transport properties of $A$SnO$_3$. We investigated the electron/thermal transport properties of Ba$_{1-x}$Sr$_x$SnO$_3$ solid solution films and their relationship with the lattice distortion. The results suggest that the all transport properties of Ba$_{1-x}$Sr$_x$SnO$_3$ are dominated by the lattice distortion. This phenomenon is attributed to the distortions in the SnO$_6$ octahedron, which consists the conduction band.

I. INTRODUCTION

Transport properties of perovskite stannate ($A$SnO$_3$, $A$ = Ba and Sr) are of great interest in advanced transparent oxide semiconductor technologies.[1,2] The edge-sharing SnO$_6$ octahedra provides a highly dispersive conduction band consists of Sn 5s orbitals with a small carrier effective mass ($m^*$) [2,3], and the $A$-site elements can change the optical bandgap[4]. For these reasons, the optoelectronic properties of $A$SnO$_3$ are highly versatile. For example, n-type BaSnO$_3$ has a cubic structure with an effective mass of ~0.4 $m_e$ and optical bandgap of ~3.2 eV.[3-5] Its single crystal exhibits an excellent carrier electron mobility ($\mu$) of ~320 cm$^{-2}$ V$^{-1}$ s$^{-1}$ at room temperature, which is comparable to classical semiconductors.[1] On the other hand, if Ba is replaced with Sr (SrSnO$_3$), the effective mass reduces to ~0.2 $m_e$ while the optical bandgap increases to ~4.6 eV.[6] This bandgap can sufficiently transmit deep-UV light (> 4.1 eV), which is crucial for DNA sensing.[7] However, despite the smaller $m^*$, the carrier electron mobility of n-type SrSnO$_3$ are significantly lower than those of n-type BaSnO$_3$ [6,8-12], which is interesting since the conduction bands of both these materials consist of SnO$_6$ octahedra.

The suppression of carrier electron mobility observed from BaSnO$_3$ ($\mu$ ~320 cm$^{-2}$ V$^{-1}$ s$^{-1}$) → SrSnO$_3$ ($\mu$ ~60 cm$^{-2}$ V$^{-1}$ s$^{-1}$) [6,12] implies that substituting Sr for Ba greatly reduces the mean free path of electrons in $A$SnO$_3$. The main scattering mechanism remains unclear but is likely related to the crystallographic transformation from cubic (BaSnO$_3$[13]) to orthorhombic (SrSnO$_3$[14]) structure (**Fig. 1**), as the SnO$_6$ octahedron will be severely distorted. This will also strongly affect vibrational properties such as thermal conductivity. In this regard, clarifying the effect of substituting Sr for Ba in BaSnO$_3$ on the evolution of both electron and thermal transport properties as well as lattice distortion is crucial for understanding the physical properties of $A$SnO$_3$ and controlling their properties for device applications.

In this study, we investigated the effect of lattice distortion on the electron and thermal transport properties of Ba$_{1-x}$Sr$_x$SnO$_3$ solid solution films. Undoped Ba$_{1-x}$Sr$_x$SnO$_3$ solid solution films were used for thermal conductivity measurements, whereas 3 % La-doped solid solution (La$_{0.03}$(Ba$_{1-x}$Sr$_x$)$_{0.97}$SnO$_3$) films were used for electron transport property measurements. Rather thick films (thickness: ~300 nm) were used to relax the strains from the lattice mismatch at the film/substrate interface.[15] The Sr content $x$ in the solid solution was varied from 0 to 1, and the data analysis was performed with a strong emphasis on establishing a relationship between the transport properties and lattice distortion. Crystallographic characterizations showed that the substitution of $A$-site ion strongly increases the deviation from a cubic structure. This overwhelms the dopant scatterings from the $A$-site substitution and dominates the changes in the electron transport properties. Thermal conductivity usually decreases with increasing atomic mass.[16] However, although Sr (87.62 amu) is lighter than Ba (137.33 amu), the thermal conductivity of $SrSnO_3$ was much lower than that of $BaSnO_3$, confirming all transport properties are dominated by the deviation from a cubic phase. These contributions reveal a fundamental factor that affects the transport properties in $ASnO_3$, which are of significant importance for engineering their transport properties in various applications.

## II.   EXPERIMENTAL

Undoped and 3 % La-doped $Ba_{1-x}Sr_xSnO_3$ solid solution films were heteroepitaxially grown on $SrTiO_3$ (001) substrates using pulsed laser deposition (KrF excimer laser, ~2 J cm$^{-2}$ pulse$^{-1}$, 10 Hz). The substrate temperature and oxygen pressure were kept at 800 °C and 10 Pa during the film growth, respectively. The thicknesses of all films were ~300 nm, which relaxes the ~~misfit~~ strain coming from the lattice mismatch between the film and the substrate. Out-of-plane X-ray Bragg diffraction patterns and reciprocal space mappings (RSMs) were measured with high-resolution X-ray diffraction (XRD, Cu K$\alpha_1$, ATX-G, Rigaku Co.) equipment. The lateral grain size ($D$) of the films was evaluated from the coherence lengths of the diffraction peaks in the RSMs. The powder XRD patterns of the solid solution targets were also measured, and the XRD patterns were analyzed with the Rietveld refinement using Rietan 2000 program [17].

The thermal conductivity ($\kappa$) of the undoped Ba$_{1-x}$Sr$_x$SnO$_3$ films in the out-of-plane direction was measured using time-domain thermo-reflectance (TDTR, PicoTR, PicoTherm Co.) at room temperature. To prevent the surface roughness from affecting the TDTR signal, the films were annealed in air at 1400 °C for ~15 min to obtain stepped and terraced surfaces (**Supplementary Fig. S1**). The electrical conductivity ($\sigma$), Hall mobility ($\mu_{Hall}$), and carrier concentration ($n$) of the La-doped La$_{0.03}$(Ba$_{1-x}$Sr$_x$)$_{0.97}$SnO$_3$ films were measured in the in-plane direction at room temperature. The Hall voltage was measured using the conventional d.c. 4-probe method with van der Pauw electrode configuration. The thermopower ($S$) of the films was measured in the in-plane direction by creating a temperature difference between two edges of the film. Details on our $S$ measurement is available elsewhere [4].

## III.  RESULTS AND DISCUSSION

**Figure 2** summarizes the XRD crystallographic analyses. Only intense diffraction peaks of 00$l_{pc}$ of Ba$_{1-x}$Sr$_x$SnO$_3$ solid solution films are seen together with 00$l$ SrTiO$_3$ substrate in the out-of-plane XRD patterns ($x$ = 0, 0.3, 0.6, and 0.9) (**Fig. 2a**). Note that the XRD characterizations of orthorhombic structure (i.e. SrSnO$_3$) represents a pseudo-cubic unit cell consists of basis atoms. For example, the pseudo-cubic unit cell of SrSnO$_3$ has Sr × 1, Sn × 1, and O × 3. The scattering vector shifts to the right with increasing $x$, which indicates a reduction in the pseudo-cubic lattice parameter in the out-of-plane direction. In order to measure the film thickness, we used Pendellösung fringes (**Fig. 2b**). **Figure 2c** shows typical RSM of the Ba$_{0.7}$Sr$_{0.3}$SnO$_3$ solid solution film around (-103) SrTiO$_3$. The lattice parameter in the out-of-plane direction ($c$-axis) was 0.41043 nm and that in the in-plane direction ($a$-axis) was 0.40853 nm, respectively. The lateral grain size ($D$) can be calculated as (integral width)$^{-1}$ in the in-plane Bragg diffraction pattern. The RSMs of all films can be found in **Supplementary Figs. S2** and **S3**.

**Figure 3** shows changes in $\kappa$ of the undoped Ba$_{1-x}$Sr$_x$SnO$_3$ solid solution films as a function of $x$. At $x = 0$ (BaSnO$_3$), $\kappa$ is the highest (8 W m$^{-1}$ K$^{-1}$). This value monotonically decreases to ~2.8 W m$^{-1}$ K$^{-1}$ at $x = 0.7$. As $x$ approaches 1, the $\kappa$ increases to 4.4 W m$^{-1}$ K$^{-1}$. The $\kappa$ of Ba$_{1-x}$Sr$_x$SnO$_3$ solid solution films exhibit a concavity and goes through a phase transition from BaSnO$_3$ dominated phase to SrSnO$_3$ dominated phase somewhere near $x$ ~0.7. This type of behavior is commonly observed in solid solution systems [18-20] due to the scattering from dopants, but the $\kappa$ of Ba$_{1-x}$Sr$_x$SnO$_3$ system is a bit different from that of conventional solid solutions. In conventional material systems, heavier atoms usually suppress lattice vibrations and decrease $\kappa$.[16] However, the $\kappa$ of SrSnO$_3$ (Sr = 87.62 amu) is lower than that of BaSnO$_3$ (Ba = 137.33 amu). This suggests that the $\kappa$ of Ba$_{1-x}$Sr$_x$SnO$_3$ solid solution films is dominated by a different factor.

**Figure 4** summarizes the electron transport properties of 3 % La-doped Ba$_{1-x}$Sr$_x$SnO$_3$ solid solution films as a function of $x$. The $\mu_{Hall}$ initially decreases from $x = 0$ (59 cm$^2$ V$^{-1}$ s$^{-1}$) to $x = 0.2$ (29 cm$^2$ V$^{-1}$ s$^{-1}$) and plateaus until $x = 0.4$ (**Fig. 4b**). Then the $\mu_{Hall}$ suddenly drops at $x = 0.5$ (12 cm$^2$ V$^{-1}$ s$^{-1}$) and slowly decreases until $x = 0.8$ (6 cm$^2$ V$^{-1}$ s$^{-1}$). As $x$ approaches from 0.8 to 1, the $\mu_{Hall}$ slowly increases up to 18 cm$^2$ V$^{-1}$ s$^{-1}$. The other electrical transport properties ($\sigma$, $n$) exhibit similar behaviors (**Figs. 4a** and **4c**), and magnitude of the thermopower ($S$) increases with decreasing $n$ (**Fig. 4c**). While the $A$-site substitution increases dopant scatterings, the changes observed in the electron transport properties are too versatile to be explained by the dopant scattering alone. The electron transport properties were not proportional to the lateral grain sizes $D$ (**Fig. 4** and **Supplementary Fig. S4(b)**). This suggests that the carrier electron mobilities are not limited by the grain boundaries. Similar to thermal conductivity, this suggests that the electron transport properties of La-doped $Ba_{1-x}Sr_xSnO_3$ solid solution films are controlled by a different factor.

Since $BaSnO_3$ and $SrSnO_3$ have different crystal structure, identifying changes in the lattice parameters is also crucial. **Figure 5** shows the lattice parameters of the undoped and La-doped $Ba_{1-x}Sr_xSnO_3$ solid solution films. The dotted lines in **Fig. 5a** and **5c** are linear regressions from the lattice parameter of $BaSnO_3$ (0.4116 nm) to that of $SrSnO_3$ (0.4034 nm). The lattice parameter in the in-plane direction is defined as $a$-axis and that in the out-of-plane direction is defined as $c$-axis. Both the $a$-axis and $c$-axis lattice parameters of undoped $BaSnO_3$ film are close to the bulk value, which indicates that the strain from the film/substrate interface is relaxed as we expected. Positive bowing is seen in $c$-axis whereas negative bowing is seen in $a$-axis, showing that the $a$-axis are in compression while the $c$-axis are in tension. The lattice distortion is defined as $(c/a-1) \times 100$ (%). The lattice distortion of the films maximizes at $x = 0.7 – 0.8$.

There are several possibilities of the origin of the lattice distortion. Since the lattice mismatch between the film and the substrate decreases with $x$, the contribution of epitaxial strain would increase. However, the in-plane lattice parameter of the resultant films was always larger than that of the substrate. Therefore, we concluded that the epitaxial strain does not dominate the lattice distortion. In order to clarify the dominant origin of the lattice distortion, we performed the Rietveld analyses of the powder diffraction patterns of the $Ba_{1-x}Sr_xSnO_3$ PLD target ceramics. We found that the space group of $x \leq 0.5$ is *Pm-3m* (cubic perovskite) whereas $x \geq 0.7$ is *Pnma* (orthorhombic perovskite) (**Supplementary Fig. S5**). The crystal structures drawn from the Rietveld analysis results also demonstrate that the lattice distortion at $x = 0.7$ is much greater than that at $x = 1$ (**Supplementary Fig. S6**). These results would indicate that the lattice distortion is spontaneously introduced due to the stability of orthorhombic symmetry around $x \sim 0.7$.

Interestingly, the changes observed in the lattice distortion are similar with those observed from the electron and thermal transport properties. In the undoped $Ba_{1-x}Sr_xSnO_3$ solid solution system, $BaSnO_3$ ($x = 0$) initially exhibits a near-perfect cubic structure, where the $\kappa$ is the highest (8 W m$^{-1}$ K$^{-1}$). The lattice distortion gradually increases until $x = 0.7$, where the $\kappa$ exhibits a minimum (**Fig. 3** and **5b**). As $x$ increases further, the lattice distortion gradually reduces until $x = 1$ ($SrSnO_3$). Since the bowing of lattice parameters occurred in the $BaSnO_3 - SrSnO_3$ solid solution system like $BaTiO_3 - SrTiO_3$ system [21], the lattice distortion was maximized around $x = 0.7$, and the cubic structure is not fully recovered, which is consistent with the $\kappa$ of $SrSnO_3$ being lower than that of $BaSnO_3$. In the case of La-doped $Ba_{1-x}Sr_xSnO_3$ solid solution films, the lattice distortion has an increasing tendency until $x = 0.8$ and decrease as it approaches $x = 1$, which is consistent with all electron transport properties (**Fig. 4** and **5d**). However, unlike the lattice distortion, a plateau can be seen in $\mu_{Hall}$ and $n$ from $x = 0.2$ to 0.4. This plateau implies that the dopant scatterings from the *A*-site substitution is not significant but not trivial to understand. It seems that ~0.6 % is a critical distortion in $SnO_6$ for the electron transport properties of $ASnO_3$. For example, a sudden drop in $\mu_{Hall}$ and $n$ is observed at $x = 0.5$ when the lattice distortion reaches ~0.7 %. In addition, an abrupt jump is observed in $\mu_{Hall}$ and $n$ at $x = 0.9$ when the lattice distortion reduces to ~0.5 %. However, we would like to note that the lattice distortions from $x = 0.2$ to 0.4 are within the uncertainty (0.1 %), and the actual lattice distortion in this region may be similar. Compared to the undoped Ba$_{1-x}$Sr$_x$SnO$_3$ solid solution films, which exhibited a transition at $x = 0.7$, the transition in the La-doped Ba$_{1-x}$Sr$_x$SnO$_3$ solid solution films occurs a bit later ($x = 0.8$). This is likely attributed to the extra distortions in the La-doped Ba$_{1-x}$Sr$_x$SnO$_3$ solid solutions from the La-dopants. In fact, the La-doped Ba$_{1-x}$Sr$_x$SnO$_3$ solid solution films exhibit greater structural distortions compared to undoped Ba$_{1-x}$Sr$_x$SnO$_3$ solid solutions near $x = 0$ and $x = 1$ (**Fig. 5b** and **5d**).

Since the propagation of vibrational waves is strongly affected by anharmonicity and strains in the lattice [22], the relationship between the lattice distortion and the $\kappa$ of Ba$_{1-x}$Sr$_x$SnO$_3$ system is not surprising. While the changes in the $\kappa$ of Ba$_{1-x}$Sr$_x$SnO$_3$ solid solution films are overall consistent with the changes observed from the lattice distortions (**Fig. 2c**), dopant scatterings also seem to play a minor role. For instance, the reduction of $\kappa$ from $x = 0$ to 0.2 is likely attributed to the dopant scattering since the lattice distortion in this region is not significant (**Fig. 2c**). In other regions, the location of the lowest $\kappa$ and the highest lattice distortion ($x = 0.7$) suggests that the effect of lattice strain is stronger. Regarding the electron transport properties, the effect of lattice distortion demonstrates the relationship between anharmonicity and the electron scattering cross sections. However, the evolution of $n$ is not straightforward. Since the optical bandgap of SrSnO$_3$ (~4.6 eV) is greater than that of BaSnO$_3$ (~3.2 eV), reduction in the La-dopant activation with the substitution of Sr is expected, but the $n$ is not monotonic with the substitution of Sr. **Figures 4b** and **5d** suggest that $n$ is also controlled by the lattice distortion. This can be related to the strain-defect coupling in oxides [23-26]. The lattice distortion creates linear strains along the axis. This can promote the formation of defects, which create defect levels and suppress the activation of dopants [27]. These results demonstrate that the electron and thermal transport properties of perovskite $A$SnO$_3$ are dominated by the lattice distortion, and maintaining the symmetry of SnO$_6$ octahedron is the key for preserving their transport properties.

## IV. SUMMARY

In summary, we investigated the effect of lattice distortion on the electron and thermal transport properties of Ba$_{1-x}$Sr$_x$SnO$_3$ solid solution films. The thermal conductivity of Ba$_{1-x}$Sr$_x$SnO$_3$ monotonically decreases from $x = 0$ to 0.7, then again monotonically increases until $x = 1$. For $x \geq 0.2$, the changes in the thermal conductivity is attributed to the scatterings from Sr substitution. However, in other regions, the lattice distortion of SnO$_6$ octahedron played a greater role again. The electron transport properties exhibited a decreasing tendency from $x = 0$ to 0.8, then increased until $x = 1$. The changes in the carrier concentrations and carrier electron mobility were highly consistent with the evolution of lattice distortions whereas the scattering from $A$-site substitutions were not significant. Maintaining the symmetry of the SnO$_6$ octahedron was a crucial factor for obtaining high electron transport properties in this system. The results of this study clarify the importance of SnO$_6$ octahedron in the transport properties of $A$SnO$_3$, which will be valuable for unlocking their potential for device applications.

**SUPPLEMENTARY MATERIAL**

See the supplementary material for additional.

**ACKNOWLEDGMENTS**

This research was supported by Grants-in-Aid for Innovative Areas (19H05791) from the JSPS. H.J.C. acknowledges the support from Nippon Sheet Glass Foundation for Materials Science and Engineering. H.O. acknowledges the support by Grants-in-Aid for Scientific Research A (17H01314) from JSPS, the Asahi Glass Foundation and the Mitsubishi Foundation. A part of this work was supported by the Dynamic Alliance for Open Innovation Bridging Human, Environment and Materials as well as the Network Joint Research Center for Materials and Devices. Student aids from Asahi glass foundation (G. Kim) and China Scholarships Council (M. Wei, 201808050081) are also greatly appreciated.

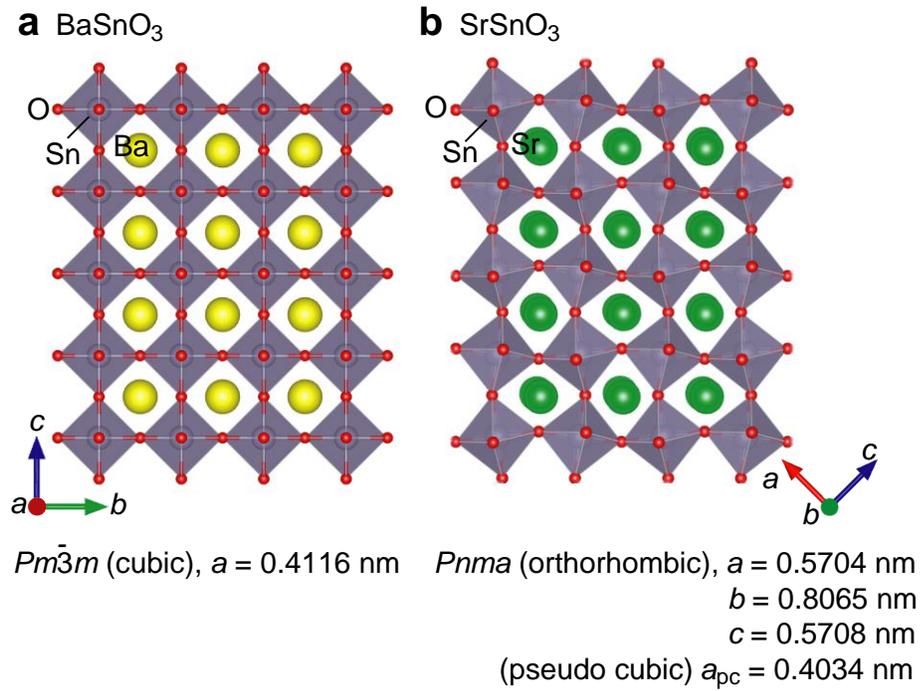

**FIG. 1** | Schematic crystal structure of (a) BaSnO$_3$ and (b) SrSnO$_3$. SnO$_6$ octahedra are aligned in the case of BaSnO$_3$ but those of SrSnO$_3$ are not. The lattice parameter of cubic BaSnO$_3$ ($a$) is 0.4116 nm and that of pseudo-cubic SrSnO$_3$ ($a_{pc}$) is = 0.4034 nm [14].

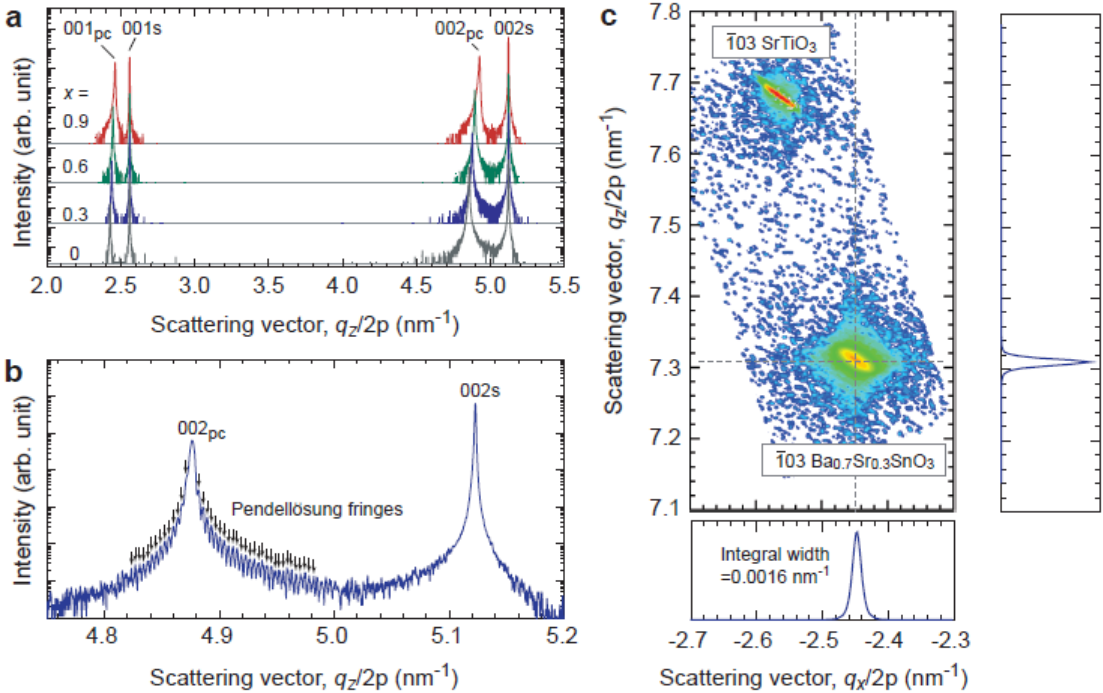

**FIG. 2 |** Crystallographic analyses. (a) Typical out-of-plane XRD patterns of undoped $Ba_{1-x}Sr_xSnO_3$ solid solution films ($x = 0$, 0.3, 0.6, and 0.9). Only intense diffraction peaks of $00l_{pc}$ of $Ba_{1-x}Sr_xSnO_3$ solid solution are seen together with $00l$ $SrTiO_3$ substrate. The scattering vector shifts to the right with increasing $x$, which indicates a reduction in the pseudo-cubic lattice parameter in the out-of-plane direction. (b) Magnified XRD pattern of the $Ba_{0.7}Sr_{0.3}SnO_3$ solid solution film. Pendellösung fringes are clearly seen. The film thickness can be measured using the Pendellösung fringes. (c) X-ray reciprocal space mapping (RSM) of the $Ba_{0.7}Sr_{0.3}SnO_3$ solid solution film around 103 $SrTiO_3$. The lattice parameter in the out-of-plane direction ($c$-axis) is 0.41043 nm and that in the in-plane direction ($a$-axis) is 0.40853 nm, respectively. The lateral grain size ($D$) can be calculated as (integral width)$^{-1}$ in the in-plane Bragg diffraction pattern.

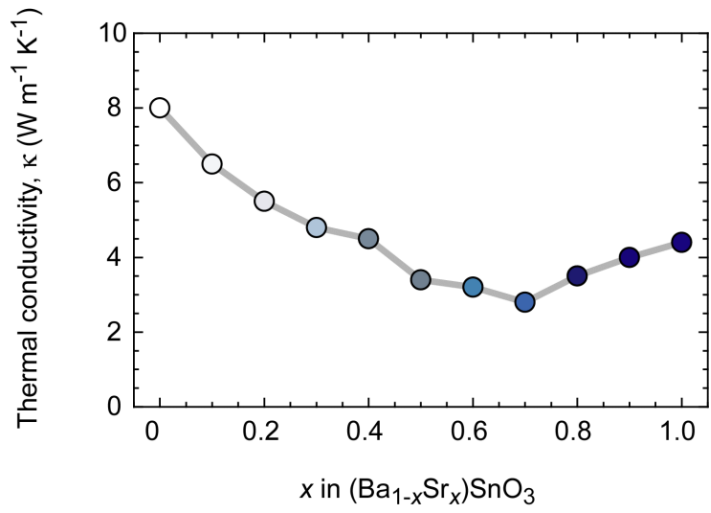

**FIG. 3** | Room temperature thermal conductivity of $Ba_{1-x}Sr_xSnO_3$ solid solution films in the out-of-plane direction.

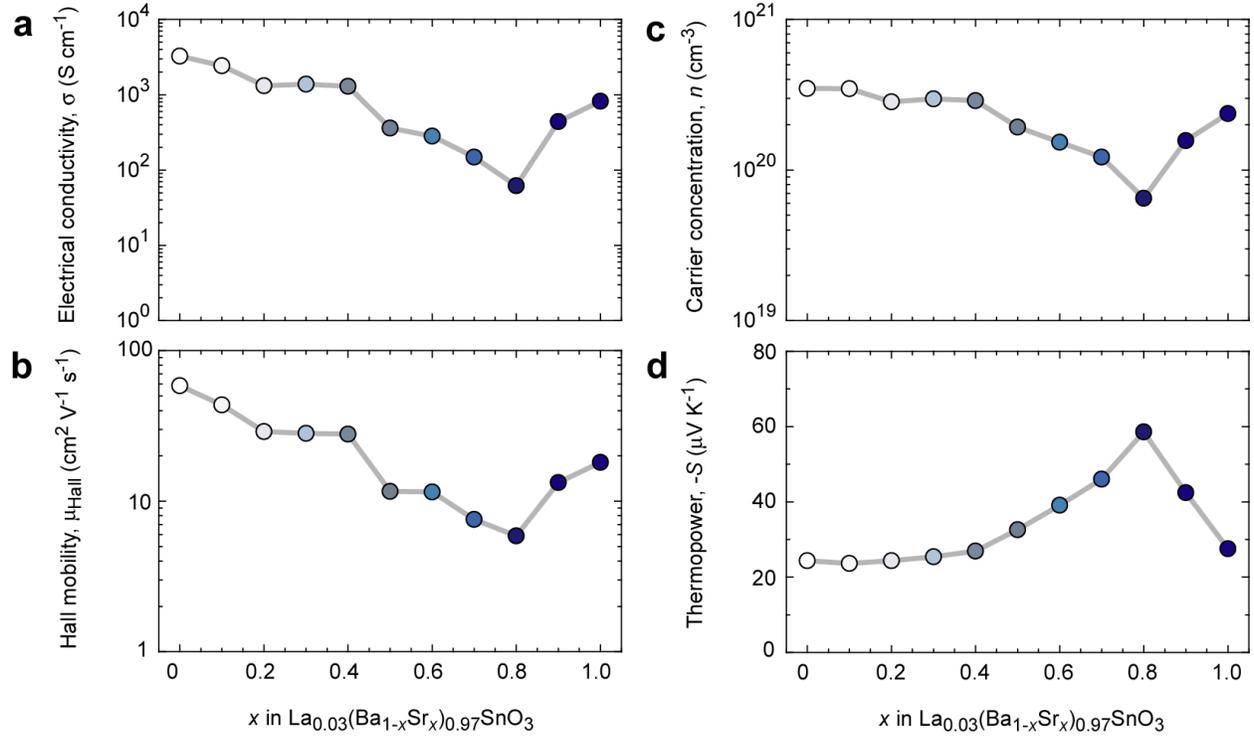

**FIG. 4** | Electron transport properties of La-doped $Ba_{1-x}Sr_xSnO_3$ solid solution films. (a) Electrical conductivity ($\sigma$), (b) Hall mobility ($\mu_{Hall}$), (c) carrier concentration ($n$) and (d) thermopower ($S$). When $x \leq 0.8$, both $\mu_{Hall}$ and $n$ decrease with $x$. Inflection points are located at $x = 0.8$ in all cases.

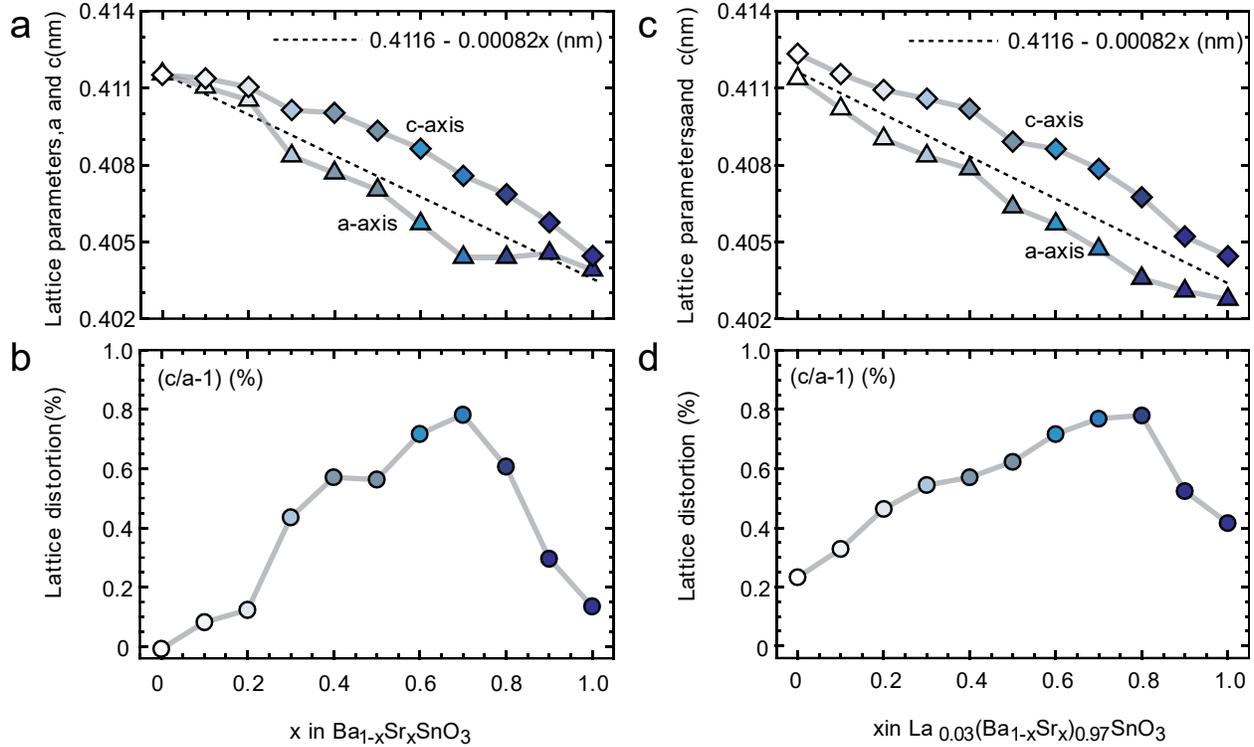

**FIG. 5** | Lattice parameters of undoped and La-doped $Ba_{1-x}Sr_xSnO_3$ solid solution films. (a) Lattice parameters and (b) lattice distortion of undoped $Ba_{1-x}Sr_xSnO_3$ films. (c) Lattice parameters and (d) lattice distortion of La-doped $Ba_{1-x}Sr_xSnO_3$ films. The dotted lines in (a) and (c) are linear regressions from the lattice parameter of $BaSnO_3$ (0.4116 nm) to that of $SrSnO_3$ (0.4034 nm). The lattice parameter in the in-plane direction is defined as $a$-axis and that in the out-of-plane direction is defined as $c$-axis. Positive bowing is seen in $c$-axis whereas negative bowing is seen in $a$-axis. The lattice distortion is defined as $(c/a-1) \times 100$ (%). The changes in the transport properties are consistent with the changes in the lattice distortions.

Supplementary Material

# Effect of lattice distortions on the electron and thermal transport properties of transparent oxide semiconductor Ba$_{1-x}$Sr$_x$SnO$_3$ solid solution films

Hai Jun Cho,[1, 2, a)] Koichi Sato,[2] Mian Wei,[2] Gowoon Kim,[2] and Hiromichi Ohta[1, 2, a)]

AFFILIATIONS

[1] Research Institute for Electronic Science, Hokkaido University, N20W10, Sapporo 001−0020, Japan

[2] Graduate School of Information Science and Engineering, Hokkaido University, N14W9, Kita, Sapporo 060−0814, Japan

[a)]Authors to whom correspondence should be addressed: joon@es.hokudai.ac.jp and hiromichi.ohta@es.hokudai.ac.jp

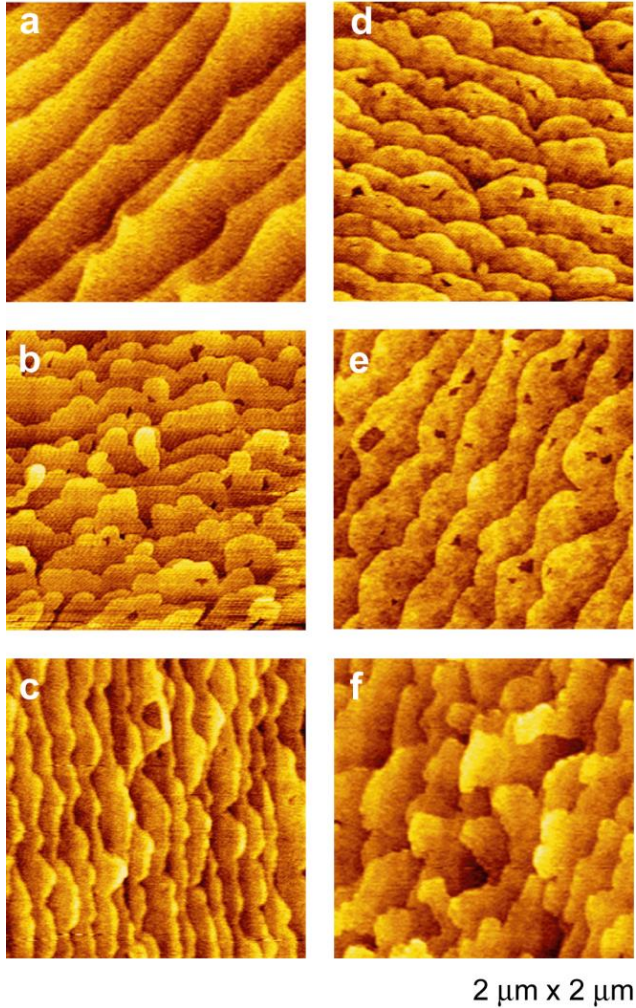

2 μm x 2 μm

**FIG. S1** | Topographic AFM images of (a) $BaSnO_3$, (b) $Ba_{0.8}Sr_{0.2}SnO_3$, (c) $Ba_{0.6}Sr_{0.4}SnO_3$, (d) $Ba_{0.4}Sr_{0.6}SnO_3$, (e) $Ba_{0.2}Sr_{0.8}SnO_3$, and (f) $SrSnO_3$ after annealing (1400 °C). Stepped and terraced surfaces are seen in all undoped solid solution films.

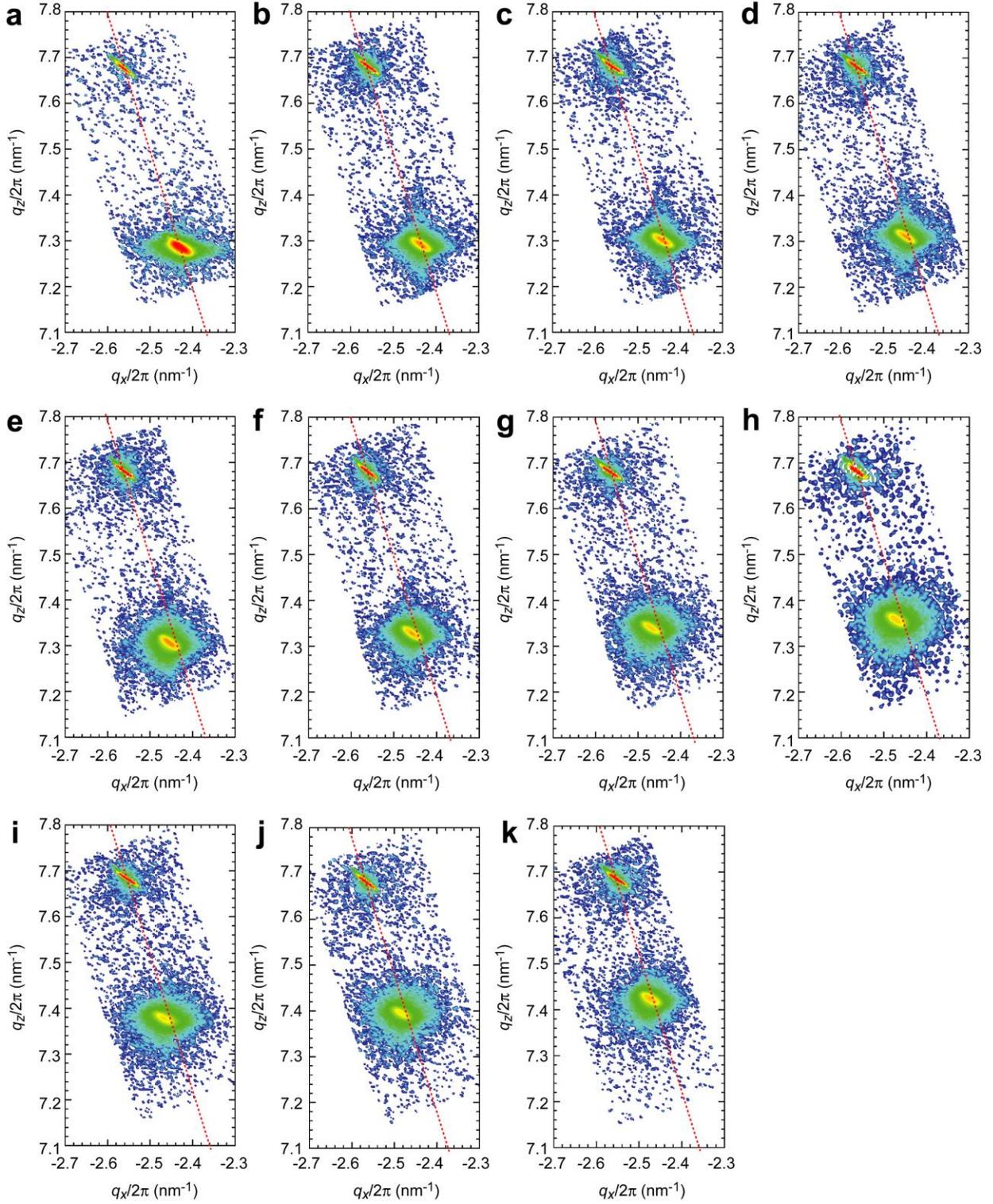

**FIG. S2 |** X-ray reciprocal space mappings (RSMs) of undoped $Ba_{1-x}Sr_xSnO_3$ solid solution films. (a) – (k): $x = 0$ to 1. Greatest deviation from the cubic line (red dotted line) occurs at $x = 0.7$.

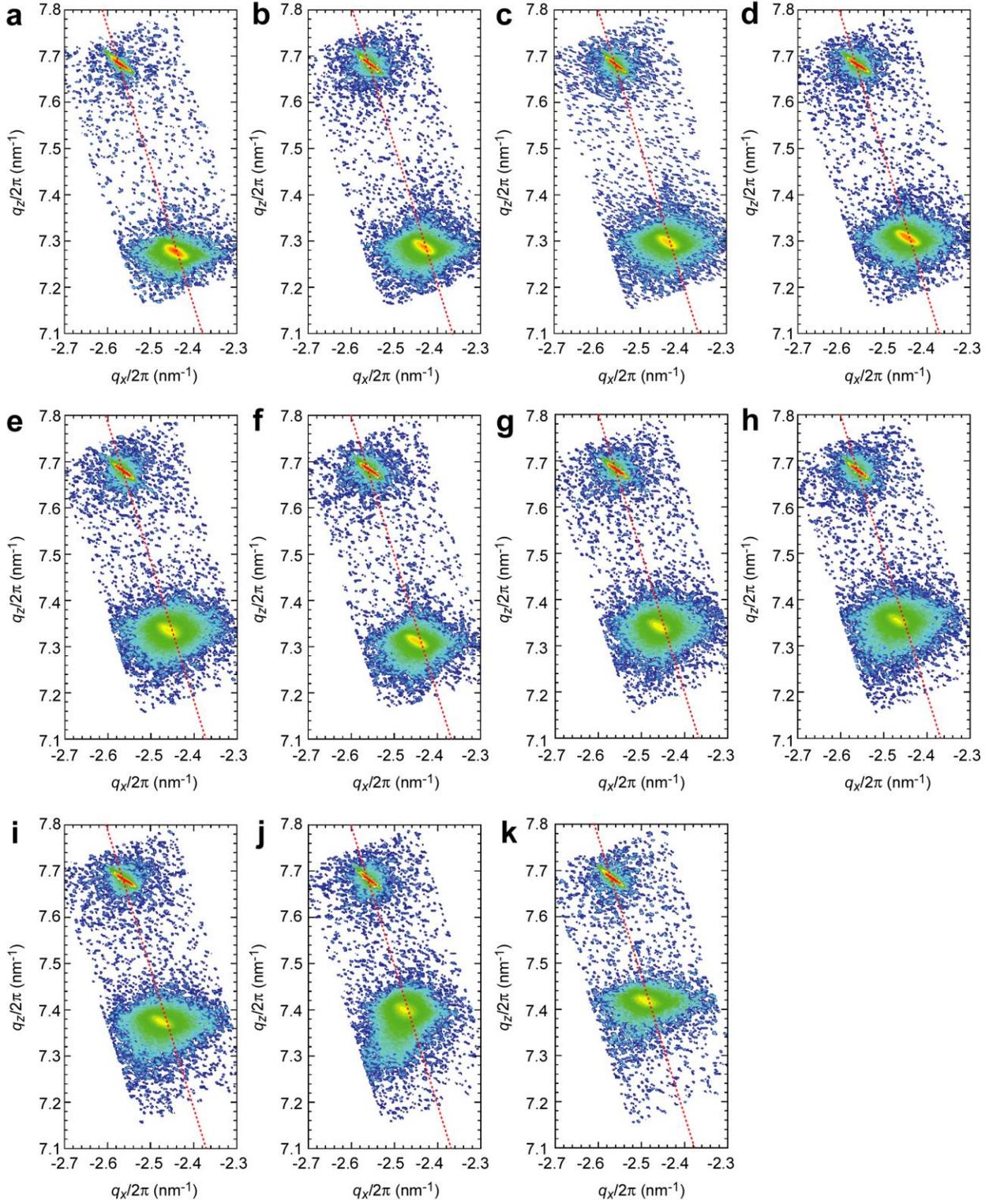

**FIG. S3 |** RSMs of La-doped Ba$_{1-x}$Sr$_x$SnO$_3$ solid solution films. (a) – (k): $x = 0$ to $1$. Greatest deviation from the cubic line (red) occurs at $x = 0.8$.

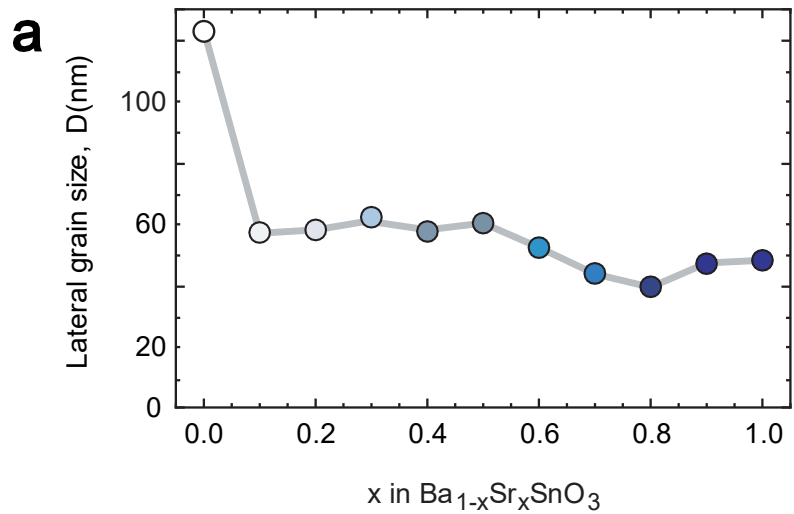

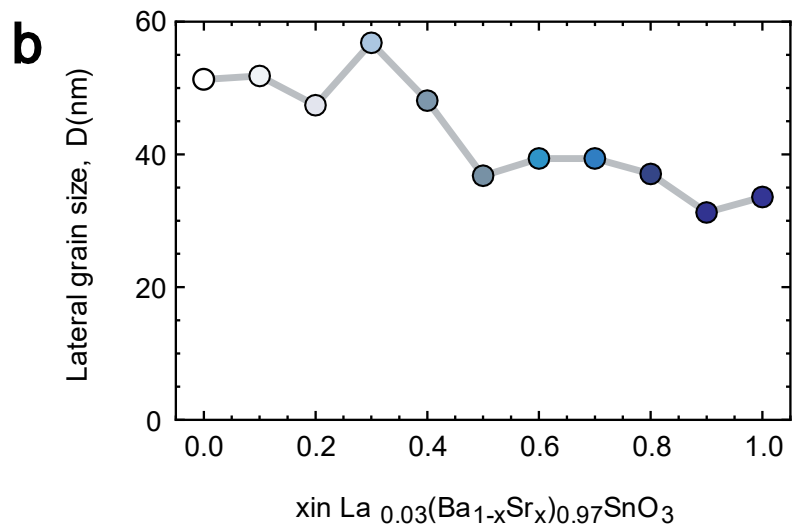

**FIG. S4 |** Lateral grain sizes of (a) undoped $Ba_{1-x}Sr_xSnO_3$ and (b) La-doped $Ba_{1-x}Sr_xSnO_3$ solid solution films.

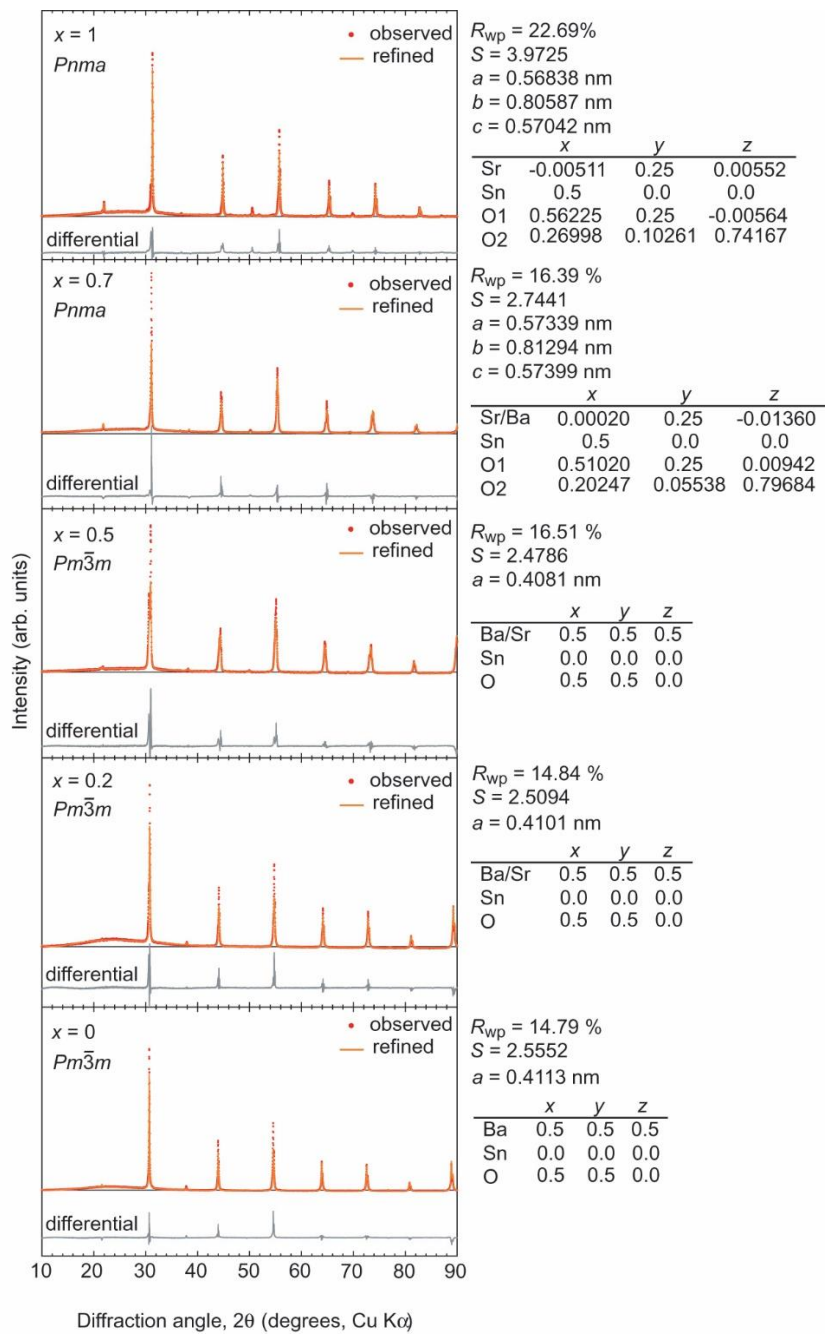

**FIG. S5** | XRD patterns of $Ba_{1-x}Sr_xSnO_3$ solid solution powders ($x$ = 0, 0.2, 0.5, 0.7 and 1) refined using Rietan 2000 program. The space group of $x$ = 0, 0.2, and 0.5 is $Pm$-$3m$ (cubic) whereas that of $x$ = 0.7 and 1 is $Pnma$ (orthorhombic).

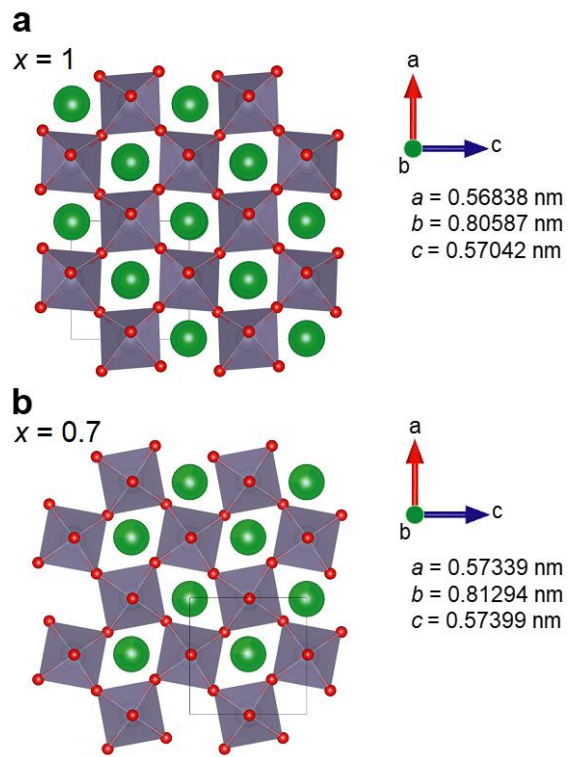

**FIG. S6** | Crystal structures of (a) SrSnO$_3$ powder and (b) Ba$_{0.3}$Sr$_{0.7}$SnO$_3$ powder. Note that the distortion of SnO$_6$ octahedra in (b) is larger than that of (a).